\documentclass[aps,prl,superscriptaddress,floatfix,twocolumn,fleqn,longbibliography]{revtex4-2}
\usepackage[utf8]{inputenc}
\usepackage{dcolumn}
\usepackage{amsbsy}
\usepackage{amsmath}
\usepackage{orcidlink}
\usepackage{amssymb}
\usepackage{braket}
\usepackage{color}
\usepackage{comment}
\usepackage{float}
\usepackage[normalem]{ulem}
\usepackage{graphicx}
\graphicspath{{./}{./Figs/}} 
\usepackage{bm}
\usepackage{hyperref}
\newcommand{\bea}{\begin{eqnarray}}
\newcommand{\eea}{\end{eqnarray}}

\def\<{\langle}
\def\>{\rangle}

\newcommand{\hide}[1]{}

\begin{document}

\title{Pinning of a High-Energy Wavepacket by a Single Impurity}
%\rmrk{Suppressing Dynamical Tunneling with a Repulsive Impurity -\small{anant}}}

\author{Fartash Chalangari \orcidlink{0009-0001-7182-204X}}
\affiliation{Computational Physics Laboratory, Tampere University, P.O. Box 600, FI-33014 Tampere, Finland}
\author{Anant Vijay Varma \orcidlink{0000-0002-7610-6317}}
\affiliation{Computational Physics Laboratory, Tampere University, P.O. Box 600, FI-33014 Tampere, Finland}
\author{Joonas Keski-Rahkonen \orcidlink{0000-0002-7906-4407}}
\affiliation{Computational Physics Laboratory, Tampere University, P.O. Box 600, FI-33014 Tampere, Finland}
\affiliation{Department of Physics, Harvard University, Cambridge, Massachusetts 02138, USA}
\affiliation{Department of Chemistry and Chemical Biology, Harvard University, Cambridge, Massachusetts 02138, USA}
\author{Esa Räsänen \orcidlink{0000-0001-8736-4496}}
\affiliation{Computational Physics Laboratory, Tampere University, P.O. Box 600, FI-33014 Tampere, Finland}
\date{\today}

\begin{abstract}
We demonstrate that a single repulsive impurity enhances directional confinement of a high-energy wavepacket in a finite two-dimensional lattice of quantum wells. At energies far above the interwell barriers, the impurity suppresses transverse dynamical tunneling while preserving propagation along a principal lattice axis, even as the corresponding classical regular region shrinks. Confinement is strongest at impurity amplitudes comparable to the barrier height and weakens for stronger perturbations. Spectral analysis links this nonmonotonic pinning to the wavepacket's overlap with eigenstates concentrated along the launch direction, consistent with impurity-induced variational scarring. These results show that a single local defect can control quantum propagation over distances far exceeding its own spatial extent.
\end{abstract}

\maketitle

%%%%%%%%%%%% Introduction %%%%%%%%%%%%%%%
\textbf{\textit{Introduction.}} Finite periodic systems provide a common setting for wavepacket transport in mesoscopic conductors, semiconductor superlattices, ultracold atoms, and photonic lattices~\cite{Beenakker1997,Datta_ElectronicTransport_book,Imry_book,Wacker2002,Bloch2008,Schwartz2007,Segev2013}. Local perturbations offer targeted control of this transport, as illustrated by experimental confinement of light through band-gap guidance at a repulsive defect~\cite{Makasyuk2006}. Controlling wavepacket propagation at energies far above the potential barriers presents a different challenge: directional confinement can depend on dynamical stability rather than energetic trapping~\cite{Graf2024,Daza2021}. This raises the question of whether a single local perturbation can suppress transverse quantum transport while preserving directional propagation in this regime.

Anderson localization and quantum scarring provide distinct mechanisms of spatial concentration. Anderson localization suppresses transport through interference among multiple scattering paths~\cite{Anderson1958,Abrahams1979,Lee1985,Evers2008,Lagendijk2009}, whereas quantum scarring enhances eigenstate probability along unstable classical periodic orbits~\cite{Heller1984,Kaplan2026}, as recently visualized in a graphene quantum dot~\cite{Ge2024,KeskiRahkonen2025}. Local perturbations can induce variational scarring by mixing nearly degenerate unperturbed states into superpositions that extremize the perturbation’s expectation value, producing enhanced probability along selected orbits of the unperturbed system~\cite{Luukko2016,KeskiRahkonen2019a}. Our previous work established the coexistence of low-energy Anderson localization and high-energy scarring along lattice channels in finite disordered arrays of quantum wells~\cite{Chalangari2026}. These channels contain a single row or column of wells, with transverse boundaries at the midpoints of the neighboring interwell barriers. While that study focused on stationary eigenstate structure and spectral statistics, here we ask whether a single local impurity can exploit this eigenstate restructuring to control high-energy wavepacket propagation.

\begin{figure}[t!]
    \centering
    \includegraphics[width=\linewidth]{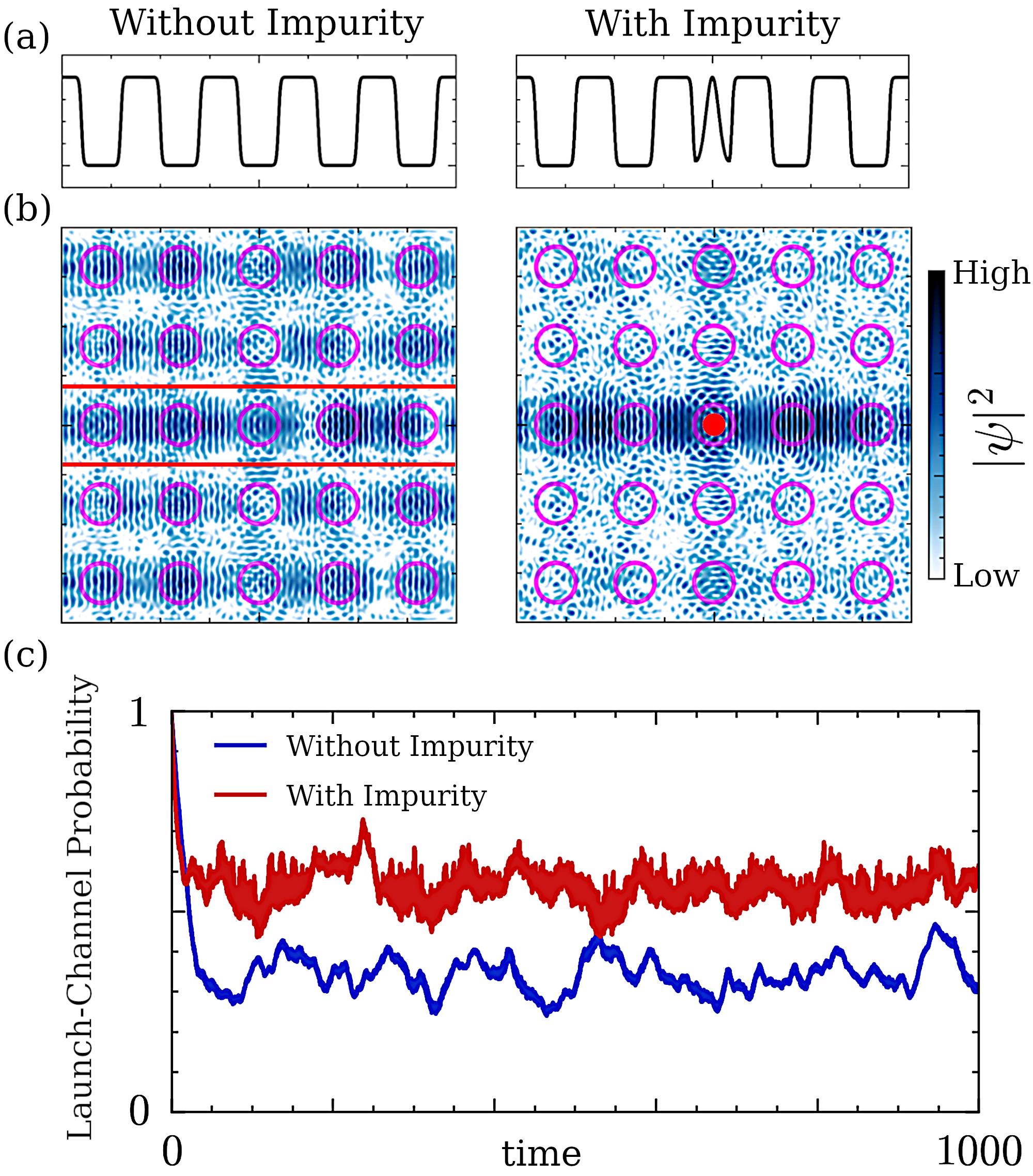}
   \caption{Pinning by a single repulsive impurity.
(a) Central-row potential profiles without (left) and with
(right) an impurity of strength $A/V_0=1$.
(b) Corresponding wavepacket probability densities at $t=808$ on a shared logarithmic scale. Magenta circles outline the wells; red lines in the left panel delimit the launch-channel and the red dot in the right panel marks the impurity.
(c) Launch-channel probability $\mathcal{P}_{\alpha_0}(t)$ for the central channel, without the impurity (blue) and with the impurity (red).}
    \label{fig:pinned_density}
\end{figure}

Here, We demonstrate that a single repulsive impurity suppresses interchannel dynamical tunneling while preserving propagation along the launch-channel in a finite square lattice of wells. Remarkably, quantum confinement strengthens even as the corresponding classical regular region shrinks. Considering a single impurity isolates the strength dependence from variations between disorder realizations. This choice is further supported by evidence that a strong local perturbation can control directional eigenstate concentration even in the presence of weaker random disorder~\cite{Selinummi2024,Chalangari2025}.

We further show that directional confinement depends nonmonotonically on impurity strength, reaching a maximum at amplitudes comparable to the interwell barrier height and weakening for stronger perturbations. Spectral analysis links this behavior to the wavepacket’s overlap with channel-concentrated eigenstates, consistent with impurity-induced variational scarring. Together, these results demonstrate how a single local defect can control high-energy wave propagation over distances far exceeding its own spatial extent.

%%%%%%%%%%%% Model Setting %%%%%%%%%%%%%%%
\textbf{\textit{Model Setting.}} We consider a single particle in a finite $5\times5$ square lattice, governed by the Hamiltonian
\begin{equation}
H=-\frac{1}{2}\nabla^2+V_{\mathrm{lattice}}(\mathbf r)
+V_{\mathrm{imp}}(\mathbf r),
\label{eq:H}
\end{equation}
with Dirichlet hard-wall boundaries and atomic units ($\hbar=m_e=e=1$). Its wavefunction obeys $i\partial_t\psi=H\psi$. The lattice comprises circular, symmetrized Fermi wells~\cite{Janecek2013},
\begin{equation}
V_{\mathrm{Fermi}}(r)
=-V_0\coth\!\left(\frac{r_w}{2d}\right)
\frac{\sinh(r_w/d)}{\cosh(r/d)+\cosh(r_w/d)},
\label{eq:fermi}
\end{equation}
where $V_0=12$ is the depth, $r_w=0.625$ the radius, and $d=0.03$ the edge softness. The lattice potential is $V_{\mathrm{lattice}}(\mathbf r)=V_0+\sum_{j=1}^{25}V_{\mathrm{Fermi}}(|\mathbf r-\mathbf r_j|)$, where $\mathbf r_j$ denotes a well center and the offset $V_0$ places the well bottoms at zero and the interwell plateau at $V_0$. The lattice constant is $a=2.5$ and the domain side length is $L=5a=12.5$. The well diameter and nominal interwell gap are equal, $2r_w=a-2r_w=1.25$.

The repulsive impurity is modeled by a Gaussian bump,
\begin{equation}
V_{\mathrm{imp}}(\mathbf r)
=A\exp\!\left[-\frac{|\mathbf r-\mathbf r_{\mathrm{imp}}|^2}
{2\sigma_{\mathrm{imp}}^2}\right],
\label{eq:Vimp}
\end{equation}
with amplitude $A\geq0$, center $\mathbf r_{\mathrm{imp}}$, and width $\sigma_{\mathrm{imp}}=0.2$. Such perturbations model smooth imperfections in quantum systems~\cite{Hirose2001,Hirose2002}. Unless stated otherwise, it lies at the central well. We vary $A/V_0$ while fixing its position, width, and all lattice parameters. Figure~\ref{fig:pinned_density}(a) compares the central-row potential profiles without and with the impurity.

The initial state is a Gaussian wavepacket,
\begin{equation}
\psi_0(\mathbf r)
=
\frac{1}{\sqrt{\pi}\sigma}
\exp\!\left[
-\frac{|\mathbf r-\mathbf r_0|^2}{2\sigma^2}
+i\mathbf p_0\cdot(\mathbf r-\mathbf r_0)
\right],
\label{eq:initial_wavepacket}
\end{equation}
centered at $\mathbf r_0$ in the central well, with width parameter $\sigma=0.5$ and mean momentum $\mathbf p_0=(16,0)$ directed along the central horizontal channel. This placement provides equivalent neighboring channels on both sides and avoids an initial transverse bias from proximity to a boundary. A matched classical ensemble is initialized from the corresponding Gaussian phase-space distribution. Writing the Hamiltonian as $H(A/V_0)$ to indicate its dependence on impurity strength, we define the initial mean energy as $E_0(A/V_0)=\langle\psi_0|H(A/V_0)|\psi_0\rangle$. In the disorder-free lattice, this equals the classical ensemble mean $\langle(p_x^2+p_y^2)/2+V_{\mathrm{lattice}}(\mathbf r)\rangle_{\mathrm{cl}}$, giving $E_0(0)\simeq132.6\simeq11V_0$. For the fixed initial state, the mean energy increases to $E_0(1)\simeq135.5$ at $A/V_0=1$, but remains conserved during each propagation. Both evolutions are followed up to $t_{\max}=1000$.

To quantify transverse confinement, we partition the domain into $M=5$ horizontal channels $\Omega_\alpha$ of width $a$. The corresponding projectors $\hat P_\alpha=\int_{\Omega_\alpha}
|\mathbf r\rangle\langle\mathbf r|\,d^2r$ satisfy
$\sum_{\alpha=1}^{M}\hat P_\alpha=\mathbb I$. The probability in channel $\alpha$ is
\begin{equation}
\mathcal P_\alpha(t)
=
\langle\psi(t)|\hat P_\alpha|\psi(t)\rangle
=
\int_{\Omega_\alpha}|\psi(\mathbf r,t)|^2\,d^2r.
\label{eq:channel_probability}
\end{equation}
The classical counterpart is the fraction of ensemble trajectories in that channel. The complete parameter set, initialization, propagation methods, and phase-space constructions are given in the Supplemental Material (SM).

\begin{figure}[t!]
    \centering
    \includegraphics[width=\linewidth]{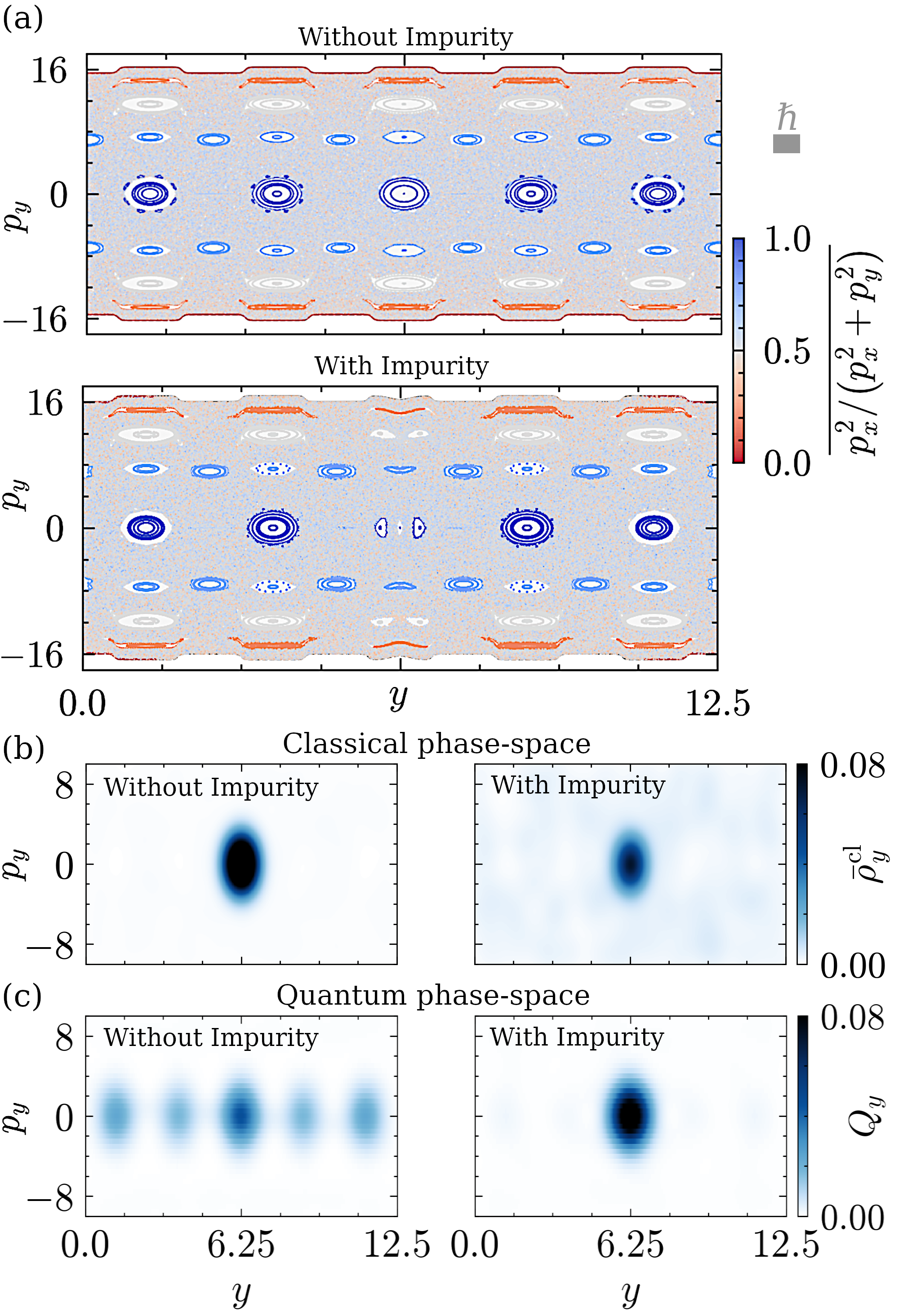}
    \caption{Classical and quantum phase-space signatures of channel confinement. (a) Poincar\'e sections in $(y,p_y)$ at $E=132.6$ without (upper) and at $E=135.5$ with (lower) an impurity of strength $A/V_0=1$. Color indicates the trajectory-averaged kinetic-energy fraction along $x$; dark-blue islands near $p_y=0$ support stable horizontal channel motion. (b) Smoothed classical densities and (c) reduced quantum Husimi distributions at $t=808$, without (left) and with (right) the impurity. At the top right, a phase-space area of $\hbar=1$ is presented.}
    \label{fig:phase-space}
\end{figure}

Figure~\ref{fig:pinned_density}(b) compares the wavepacket densities at a representative late time, $t=808$, with red lines delimiting the launch-channel in the left panel. In the disorder-free lattice (left), probability has transferred into the other four channels. A central impurity with $A/V_0=1$ strongly reduces this transfer; the wavepacket remains predominantly concentrated in the launch-channel, with a weaker enhancement along the intersecting vertical column and a diffuse background (right). The diffuse background is consistent with the wavepacket’s finite spatial and momentum spreads sampling regions outside the regular channel islands. The launch-channel probability confirms enhanced confinement throughout much of the evolution [Fig.~\ref{fig:pinned_density}(c)]. We refer to this enhanced transverse confinement, accompanied by continued longitudinal motion, as \emph{pinning}.

%%%%%%%%%%% Dynamical Tunneling %%%%%%%%%%%%%%
\textbf{\textit{Dynamical Tunneling.}} To understand this behavior, we compare the classical and quantum phase-space dynamics in Fig.~\ref{fig:phase-space}. The Poincar\'e sections pass through the lattice center at the respective initial mean energies, $E_0(0)=132.6$ and $E_0(1)=135.5$, with color indicating the trajectory-averaged fraction of kinetic energy along $x$. The disorder-free section [Fig.~\ref{fig:phase-space}(a), upper panel] exhibits mixed phase space, with dark-blue regular islands supporting stable horizontal-channel motion. Panels (b) and (c) show the smoothed classical density $\bar{\rho}_y^{\mathrm{cl}}(y,p_y,t)$ and reduced quantum Husimi function $Q_y(y,p_y,t)$, respectively (definitions in the SM). At $t=808$, the classical distribution remains concentrated in the launch-channel [panel (b), left], whereas the quantum Husimi distribution exhibits five concentrations associated with the five channels [panel (c), left].

This contrast indicates dynamical tunneling: quantum transfer between regions separated by classical phase-space barriers rather than potential-energy barriers~\cite{Davis1981,Bohigas1993,Graf2024}. Regular classical trajectories remain confined by the invariant structures bounding each channel region, whereas quantum probability can cross between them.

%%%%%%%%%%% Pinning %%%%%%%%%%%%%
\textbf{\textit{Pinning.}} The impurity splits and reduces the central regular island structure [Fig.~\ref{fig:phase-space}(a), lower], consistent with a bifurcation of the underlying periodic orbit. The classical ensemble correspondingly becomes less concentrated in the launch-channel and spreads more widely [Fig.~\ref{fig:phase-space}(b), right]. The classical launch-channel probability confirms reduced confinement for $A/V_0=1$ and $4$ relative to the disorder-free case after the initial transient (see Fig.~\ref{fig:classical_channel_probability}). Because the ensemble samples the initial Gaussian phase-space distribution, this comparison incorporates the launch energy spread.

In contrast, the quantum Husimi distribution develops a dominant central concentration as neighboring-channel populations diminish [Fig.~\ref{fig:phase-space}(c), right], resembling the confined disorder-free classical distribution. Enhanced quantum confinement despite the smaller classical regular region and increased classical spreading supports its interpretation as suppression of interchannel dynamical tunneling.

The central impurity preserves the lattice's $C_{4v}$ symmetry, leaving horizontal and vertical orientations equivalent. The initial wavepacket's horizontal momentum favors overlaps with eigenstates supporting horizontal motion, while column support in the contributing eigenstates and their interference can account for the weaker vertical enhancement. This persistent directional preference is reminiscent of quantum birthmarks, which describe lasting memory of the initial state and its early dynamics~\cite{Graf2026}. Off-center impurities yield comparable confinement at the same amplitude and width (see Fig.~\ref{fig:off-center}), supporting the central placement as representative.

\begin{figure}[th!]
    \centering
    \includegraphics[width=\linewidth]{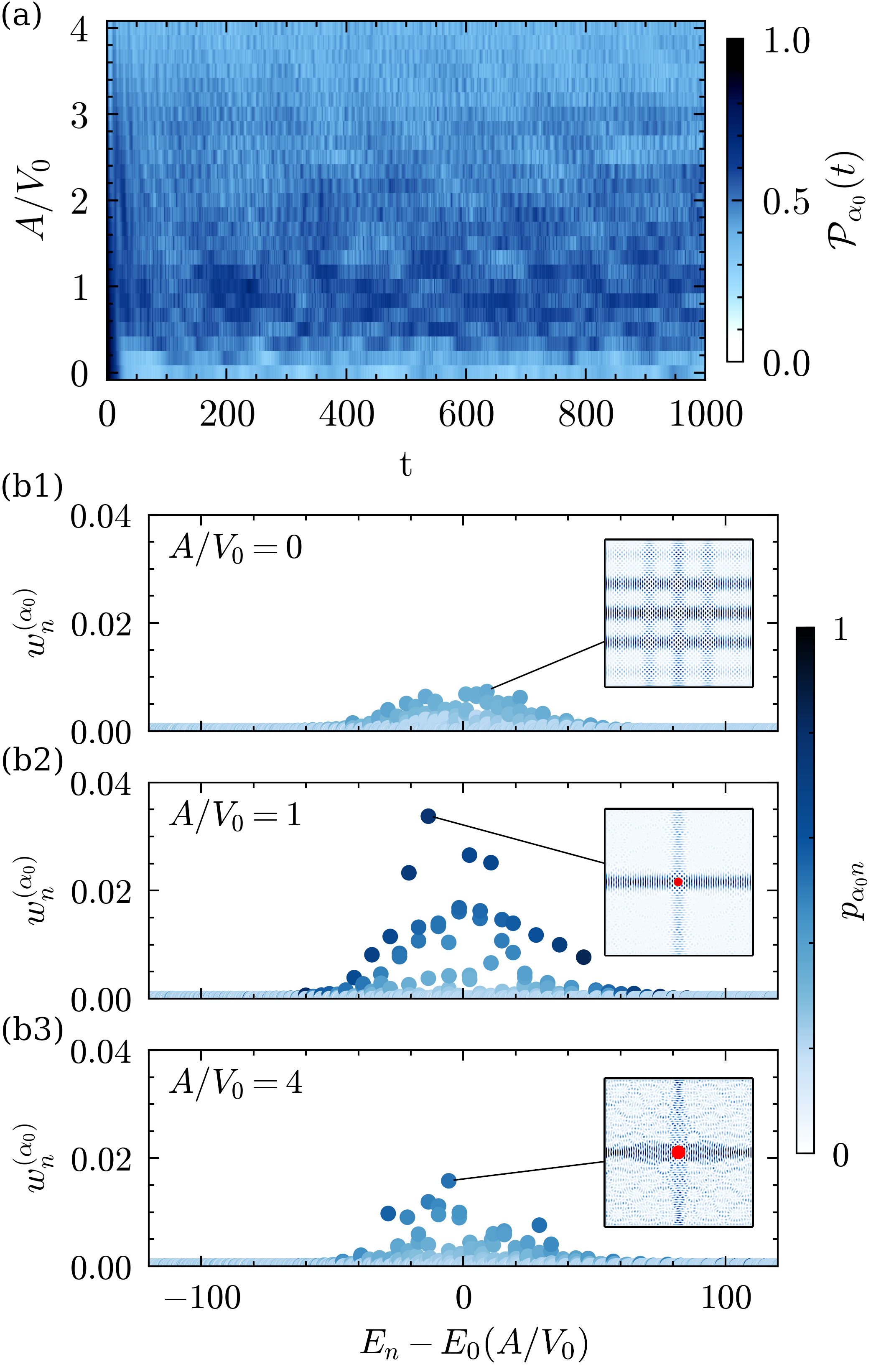}
    \caption{Nonmonotonic channel confinement and its spectral support. (a) Launch-channel probability $\mathcal P_{\alpha_0}(t)$ versus time and impurity strength $A/V_0$, with impurity position and width fixed. (b1)--(b3) Diagonal channel weights $w_n^{(\alpha_0)}$ for $A/V_0=0,1,$ and $4$, respectively, versus $E_n-E_0$, where $E_0(A/V_0)$ is the initial wavepacket's mean energy for each impurity strength. Color indicates the eigenstate launch-channel probability $p_{\alpha_0n}$. Insets show the density of the eigenstate with the largest weight; red dots mark the impurity position.}
    \label{fig:spectral_comparison}
\end{figure}

\textbf{\textit{Role of Scarring.}} Variational scarring provides a framework for understanding how a local impurity can favor eigenstates concentrated along an extended channel. In this picture, the impurity mixes degenerate or nearly degenerate unperturbed states associated with classical resonances. Diagonalizing the impurity potential within this subspace identifies combinations that extremize its expectation value. These combinations approximate the perturbed eigenstates when impurity-induced mixing dominates the energy splittings and contributions from states outside the subspace remain small. Such combinations can exhibit enhanced probability density along selected classical periodic orbits of the unperturbed system~\cite{Luukko2016,KeskiRahkonen2017,KeskiRahkonen2019b}, but their contribution to wavepacket confinement also requires appreciable overlap with the initial state. We therefore examine channel concentration and wavepacket overlap together to identify the spectral support for confinement.

With impurity position and width fixed, confinement is strongest at $A/V_0\sim1$ and weakens for stronger impurities [Fig.~\ref{fig:spectral_comparison}(a)]. To examine this nonmonotonic behavior, we expand the wavepacket in $N$ eigenstates of the Hamiltonian in Eq.~\eqref{eq:H}, spanning an energy range around the initial wavepacket’s mean energy,
\begin{equation}
|\psi_{\alpha_0}^{(N)}(t)\rangle
=\sum_{n=1}^{N}c_n^{(\alpha_0)}e^{-iE_nt}|\phi_n\rangle,
\label{eq:retained_spectral_state}
\end{equation}
where $H|\phi_n\rangle=E_n|\phi_n\rangle$ and $c_n^{(\alpha_0)}=\langle\phi_n|\psi_{\alpha_0}(0)\rangle$. We choose $N$ sufficiently large that $\sum_{n=1}^{N}|c_n^{(\alpha_0)}|^2\simeq1$, retaining essentially the full wavepacket norm. Here, $\alpha_0$ denotes the launch-channel and $\alpha$ the observation channel. The channel projector matrix elements are $P_{nm}^{(\alpha)}=\langle\phi_n|\hat P_\alpha|\phi_m\rangle$, with $p_{\alpha n}=P_{nn}^{(\alpha)}$ giving the eigenstate probability in channel $\alpha$.

We define the diagonal launch-channel weights as
\begin{equation}
w_n^{(\alpha_0)}
\equiv |c_n^{(\alpha_0)}|^2 \, p_{\alpha_0n},
\label{eq:eigenstate_pin_weight}
\end{equation}
which are substantial only when appreciable packet overlap coincides with strong launch-channel support. Their sum gives the time-independent diagonal contribution,
\begin{equation}
\mathcal P_{\alpha_0}^{\mathrm{diag}}
=\sum_{n=1}^{N}w_n^{(\alpha_0)}.
\label{eq:diagonal_population_contributions}
\end{equation}
Within this expansion,
\begin{equation}
\mathcal P_\alpha(t)
=\mathcal P_\alpha^{\mathrm{diag}}+\mathcal C_\alpha(t),
\label{eq:channel_probability_decomposition}
\end{equation}
where the off-diagonal interference contribution is
\begin{equation}
\mathcal C_\alpha(t)
=\sum_{\substack{n,m=1\\n\neq m}}^{N}
c_n^{(\alpha_0)*}c_m^{(\alpha_0)}
P_{nm}^{(\alpha)}e^{i(E_n-E_m)t}.
\label{eq:channel_population_components}
\end{equation}
Thus, the diagonal term measures packet-weighted eigenstate channel support, while interference $\mathcal C_\alpha(t)$ determines deviations from it. Energy differences set the oscillation frequencies; exactly degenerate pairs contribute time-independent terms.

Figures~\ref{fig:spectral_comparison}(b1)--(b3) show the diagonal channel weights versus $E_n-E_0$, with $E_0$ evaluated for each impurity strength. The horizontal launch preferentially samples eigenstates with horizontal structure. In the disorder-free lattice [Fig.~\ref{fig:spectral_comparison}(b1)], strongly contributing states exhibit stripes across several channels. Their interference initially concentrates probability in the launch-channel, while evolving relative phases redistribute it among channels, providing spectral support for dynamical tunneling. At $A/V_0=1$ [Fig.~\ref{fig:spectral_comparison}(b2)], appreciable packet overlap coincides with greater launch-channel concentration, producing larger weights. At $A/V_0=4$ [Fig.~\ref{fig:spectral_comparison}(b3)], the largest weights decrease, and strongly contributing states generally have lower launch-channel probabilities. Darker markers indicate larger $p_{\alpha_0n}$; each inset shows the state with the largest weight, illustrating enhanced channel concentration at intermediate strength and increased density outside the launch row at stronger perturbation. This reduction is consistent with the weakening of impurity-induced scarring observed in our previous study~\cite{Chalangari2026}. The confinement maximum occurs when appreciable overlap with the initial wavepacket coincides with strong eigenstate concentration in the launch-channel.

\begin{figure}[t]
    \centering
    \includegraphics[width=\linewidth]{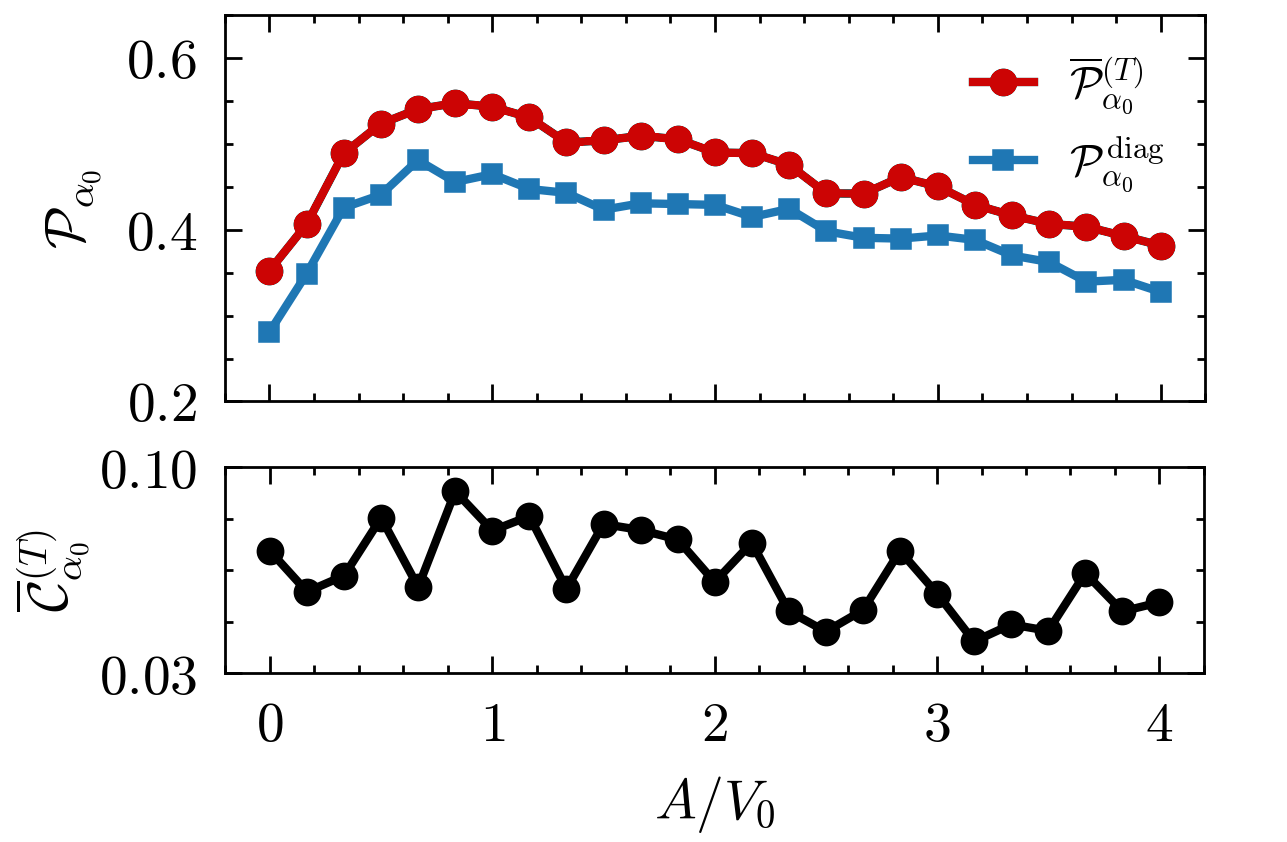}
    \caption{Diagonal and coherent contributions to launch-channel confinement. Upper: finite-time mean $\overline{\mathcal P}_{\alpha_0}^{(T)}$ over $0\leq t\leq T=1000$ and diagonal contribution $\mathcal P_{\alpha_0}^{\mathrm{diag}}$ versus impurity strength. Lower: their difference, the mean coherence contribution $\overline{\mathcal C}_{\alpha_0}^{(T)}$.}
    \label{fig:nonmonotonic}
\end{figure}

The summed diagonal contribution captures the main nonmonotonic trend [Fig.~\ref{fig:nonmonotonic}, upper]. To compare it with confinement over the observation interval, we evaluate the finite-time mean channel probability as
\begin{equation}
\begin{aligned}
\overline{\mathcal P}_\alpha^{(T)}
&\equiv\frac{1}{T}\int_0^T\mathcal P_\alpha(t)\,dt\\
&=\mathcal P_\alpha^{\mathrm{diag}}
+\sum_{\substack{n,m=1\\n\neq m}}^{N}
c_n^{(\alpha_0)*}c_m^{(\alpha_0)}P_{nm}^{(\alpha)}\\
&\quad\times e^{i(E_n-E_m)T/2}
\operatorname{sinc}\!\left[\frac{(E_n-E_m)T}{2}\right],
\end{aligned}
\label{eq:finite_time_channel_average}
\end{equation}
where $\operatorname{sinc}(z)=\sin(z)/z$ and $\operatorname{sinc}(0)=1$. This expression is derived and validated against direct split-operator propagation in the SM. It retains time-independent interference between exactly degenerate states. Within degenerate subspaces, the diagonal and interference contributions can depend on the chosen eigenbasis, but the total finite-time mean channel probability is basis independent.

For $T=1000$, the mean exceeds the diagonal contribution throughout the scan. Their difference,
$\overline{\mathcal C}_{\alpha_0}^{(T)}
=\overline{\mathcal P}_{\alpha_0}^{(T)}
-\mathcal P_{\alpha_0}^{\mathrm{diag}}$
[Fig.~\ref{fig:nonmonotonic}, lower], therefore enhances mean confinement, although its instantaneous counterpart can be negative. Small energy gaps account for most of this correction, with little further net change from larger gaps (see cumulative coherence analysis in the SM), consistent with rapidly oscillating contributions averaging out. These contributions can still produce instantaneous fluctuations; averaging does not imply loss of coherence.

Changing the launch-channel provides a complementary test. A packet launched in a channel without impurity spreads among the unperturbed channels, while the channel with impurity remains the least populated after the initial transient (Fig.~\ref{fig:diffLaunch} in the SM). Changing the initial state changes its eigenstate overlaps but leaves the Hamiltonian and channel projector matrix elements unchanged. This behavior is consistent with weak overlap with states concentrated in the impurity channel and stronger overlap with states extending across the unperturbed channels. The impurity thus enhances confinement for a packet launched in its channel while limiting population of that channel for a packet launched elsewhere.

%%%%%%%%%%%%%%%%%% Summary %%%%%%%%%%%%%%%%%%%%%
\textbf{\textit{Summary.}}
Together, these results demonstrate pinning of a high-energy wavepacket along an extended channel by a single local repulsive impurity. Confinement is strongest at intermediate impurity strength, where appreciable wavepacket overlap coincides with strong eigenstate concentration along the launch-channel. Because longitudinal propagation persists far above the interwell barriers, this pinning is dynamical rather than conventional energetic trapping. The enhancement occurs under unitary quantum evolution even as the corresponding classical regular region shrinks, showing that nonintegrable dynamics need not prevent controlled directional propagation.

Although quantitative confinement depends on geometry, our previous observation of impurity-induced scarring across different spatial scales~\cite{Chalangari2026} suggests that pinning may extend beyond the present configuration when channel-concentrated states retain appreciable overlap with the initial wavepacket. System size, launch energy and the arrangement of impurities affect confinement and its long-time evolution through eigenstate overlaps, channel probabilities and energy splittings. These findings motivate further theoretical and experimental exploration of local control of high-energy wave propagation in photonic and solid-state systems.

%%%%%%%%%%%%% Acknowledgment %%%%%%%%%%%%%%%%
\textbf{\textit{Acknowledgment.}} We acknowledge CSC---IT Center for Science, Finland, for computational resources and the Research Council of Finland, ManyBody2D Project No.~349956, for financial support. This project is also supported by the National Science Foundation (Grant No.~2403491).

\bibliography{Bibliography}

%%%%%%%%%%%%%%%%%%%%%%%%%%%%%%%%%%%%%%%%%%%%%%

\onecolumngrid
\pagestyle{empty}

\renewcommand{\thefigure}{S\arabic{figure}}
\setcounter{figure}{0}
\renewcommand{\theequation}{S-\arabic{equation}}
\setcounter{equation}{0}

\clearpage

\begin{center}
{\large\bfseries Supplemental Material for\\[0.5em]
Pinning of a High-Energy Wavepacket by a Single Impurity}

\vspace{1em}

{Fartash Chalangari, Anant Vijay Varma, Joonas Keski-Rahkonen, and Esa Räsänen}
\end{center}

\section{Quantum Wavepacket Dynamics}

The initial state is a minimum-uncertainty Gaussian wavepacket,
\begin{equation}
\psi_0(\mathbf r)
=
\frac{1}{\sqrt{\pi}\sigma}
\exp\!\left[
-\frac{|\mathbf r-\mathbf r_0|^2}{2\sigma^2}
+i\mathbf p_0\cdot(\mathbf r-\mathbf r_0)
\right],
\label{eq:initial_wavepacket_SM}
\end{equation}
with $\sigma=0.5$, center $\mathbf r_0$ in the launch well, and mean momentum $\mathbf p_0=(16,0)$~\cite{Heller_book}. Unless stated otherwise, the launch-channel is the central horizontal row, $\alpha_0=3$. Neglecting the small boundary corrections, the rms uncertainties are
\begin{equation}
\Delta x=\Delta y=\frac{\sigma}{\sqrt{2}},
\qquad
\Delta p_x=\Delta p_y=\frac{1}{\sqrt{2}\sigma}.
\label{eq:initial_uncertainties_SM}
\end{equation}
Including the momentum spread gives
\begin{equation}
\langle T(0)\rangle
=
\frac{|\mathbf p_0|^2}{2}
+\frac{1}{2\sigma^2}
=130.
\label{eq:initial_kinetic_energy_SM}
\end{equation}
The disorder-free initial mean energy in the numerical energy convention is $E_0=\langle\psi_0|H|\psi_0\rangle \simeq132.6\simeq11V_0$.

The state evolves according to
\begin{equation}
i\partial_t\psi(\mathbf r,t)=H\psi(\mathbf r,t).
\label{eq:tdse_SM}
\end{equation}
We compute its evolution using two complementary methods. Spectral propagation uses
\begin{equation}
\psi_{\alpha_0}^{(N)}(\mathbf r,t)
=
\sum_{n=1}^{N}
c_n^{(\alpha_0)}e^{-iE_nt}\phi_n(\mathbf r),
\label{eq:spectral_propagation}
\end{equation}
where $H|\phi_n\rangle=E_n|\phi_n\rangle$ and $c_n^{(\alpha_0)} =\langle\phi_n|\psi_{\alpha_0}(0)\rangle$. The $N=6400$ eigenstates, computed by imaginary-time propagation~\cite{Luukko2013}, capture approximately $99.9\%$ of the initial probability across the cases studied:
\begin{equation}
W=\sum_{n=1}^{N}|c_n^{(\alpha_0)}|^2\simeq0.999.
\label{eq:retained_norm_SM}
\end{equation}
We retain the original coefficients without renormalizing
the truncated state; its norm remains $W$ during spectral
propagation. Independently, a split-operator scheme~\cite{Hermann1988}
propagates the wavepacket on a uniform $400\times400$ spatial
grid covering the $12.5\times12.5$ domain. The kinetic
step uses a sine basis to enforce Dirichlet boundary conditions. We use a time step $\Delta t=10^{-3}$ for $10^6$ steps, reaching $t_{\max}^{\mathrm q}=1000$. Spectral propagation covers the same observation interval.

\subsection{Reduced Husimi distributions}

Directional phase-space structure is characterized by reduced Husimi functions~\cite{Husimi1940},
\begin{equation}
\begin{aligned}
Q_x(x_0,p_x,t)
&=
\frac{1}{2\pi}\int dy\,
\left|
\int dx'\,
\varphi_{x_0,p_x;\sigma}^{*}(x')
\psi(x',y,t)
\right|^2,\\
Q_y(y_0,p_y,t)
&=
\frac{1}{2\pi}\int dx\,
\left|
\int dy'\,
\varphi_{y_0,p_y;\sigma}^{*}(y')
\psi(x,y',t)
\right|^2.
\end{aligned}
\label{eq:reduced_husimi}
\end{equation}
The normalized one-dimensional Gaussian probe is
\begin{equation}
\varphi_{q_0,p_q;\sigma}(q)
=
\frac{1}{\pi^{1/4}\sqrt{\sigma}}
\exp\!\left[
-\frac{(q-q_0)^2}{2\sigma^2}
+ip_q(q-q_0)
\right].
\label{eq:husimi_probe_SM}
\end{equation}
Here, $(q_0,p_q)$ denotes the probe center. The wavefunction is taken to vanish outside the spatial domain. We fix the probe width at $\sigma=0.5$ for all times, eigenstates, and impurity strengths.

\section{Classical Dynamics and Phase-Space Comparison}

For comparison, we propagate an ensemble of $N_{\mathrm{cl}}=10\,000$ trajectories in the same potential, with specular reflection at the hard walls. Trajectories are
integrated with time step $\Delta t_{\mathrm{cl}}=10^{-3}$ up to $t_{\max}^{\mathrm{cl}}=1000$. Initial conditions are sampled from
\begin{equation}
\mathbf z=(x,y,p_x,p_y)
\sim\mathcal N(\boldsymbol\mu,\boldsymbol\Sigma),
\label{eq:classical_sampling}
\end{equation}
where $\boldsymbol\mu=(x_0,y_0,p_{x0},p_{y0})$ and
\begin{equation}
\boldsymbol\Sigma
=
\operatorname{diag}\!\left(
\frac{\sigma^2}{2},\frac{\sigma^2}{2},
\frac{1}{2\sigma^2},\frac{1}{2\sigma^2}
\right).
\label{eq:classical_covariance_SM}
\end{equation}
This distribution matches the positive Wigner function of the initial Gaussian wavepacket~\cite{Hillery1984},
\begin{equation}
W_0(\mathbf r,\mathbf p)
=
\frac{1}{\pi^2}
\exp\!\left[
-\frac{|\mathbf r-\mathbf r_0|^2}{\sigma^2}
-\sigma^2|\mathbf p-\mathbf p_0|^2
\right],
\label{eq:initial_wigner_SM}
\end{equation}
up to negligible boundary corrections. The quantum state and classical ensemble thus have matching initial position and momentum distributions and mean energies, within sampling accuracy. Classical channel probabilities are the fractions of trajectories within the corresponding channels.

\subsection{Reduced classical distributions}

The instantaneous classical phase-space density $f_{\mathrm{cl}}(x,y,p_x,p_y,t)$ is reduced to the same planes as the Husimi functions:
\begin{equation}
\begin{aligned}
f_x^{\mathrm{cl}}(x,p_x,t)
&=
\int dy\,dp_y\,
f_{\mathrm{cl}}(x,y,p_x,p_y,t),\\
f_y^{\mathrm{cl}}(y,p_y,t)
&=
\int dx\,dp_x\,
f_{\mathrm{cl}}(x,y,p_x,p_y,t).
\end{aligned}
\label{eq:reduced_classical_distributions}
\end{equation}
To compare quantum and classical distributions at the same phase-space resolution, we convolve each classical marginal with a normalized Gaussian kernel:
\begin{equation}
\begin{aligned}
\bar\rho_q^{\mathrm{cl}}(q,p_q,t)
&=
\int dq'\,dp_q'\,
G_\sigma(q-q',p_q-p_q')\\
&\qquad\times f_q^{\mathrm{cl}}(q',p_q',t),
\qquad q\in\{x,y\},
\end{aligned}
\label{eq:coarse_grained_classical}
\end{equation}
where
\begin{equation}
G_\sigma(u,v)
=
\frac{1}{\pi}
\exp\!\left(-\frac{u^2}{\sigma^2}-\sigma^2v^2\right).
\label{eq:classical_smoothing_kernel_SM}
\end{equation}
Its rms widths match the Gaussian-probe uncertainties in Eq.~\eqref{eq:husimi_probe_SM}. For display, both quantum and classical maps are normalized to unit integral over the full phase plane. This display normalization does not alter the unrenormalized spectral probabilities used in the channel analysis. These distributions are instantaneous projections, with no surface-crossing condition. They are therefore distinct from Poincar\'e sections.

\subsection{Poincar\'e sections}

We characterize the classical phase-space structure using sections through the lattice center, recording $(y,p_y)$ at the central vertical section and $(x,p_x)$ at the central horizontal section. Initial conditions are sampled on a $31\times31$ grid in the section coordinate and its conjugate momentum, giving up to 961 energetically allowed initial conditions per section. Each trajectory is integrated with time step $10^{-3}$ up to $t=1000$. We record positive-direction crossings, i.e., $x=0$ with $p_x>0$ for the $(y,p_y)$ section, and $y=0$ with $p_y>0$ for the $(x,p_x)$ section, using coordinates centered on the lattice. The sections in Fig.~\ref{fig:phase-space}(a) of the main text are evaluated at $E=132.6$ for the disorder-free lattice and $E=135.5$ for $A/V_0=1$, matching the respective initial wavepacket mean energies.

\subsection{Classical launch-channel probability}

\begin{figure}[h]
    \centering
    \includegraphics[width=0.4\linewidth]{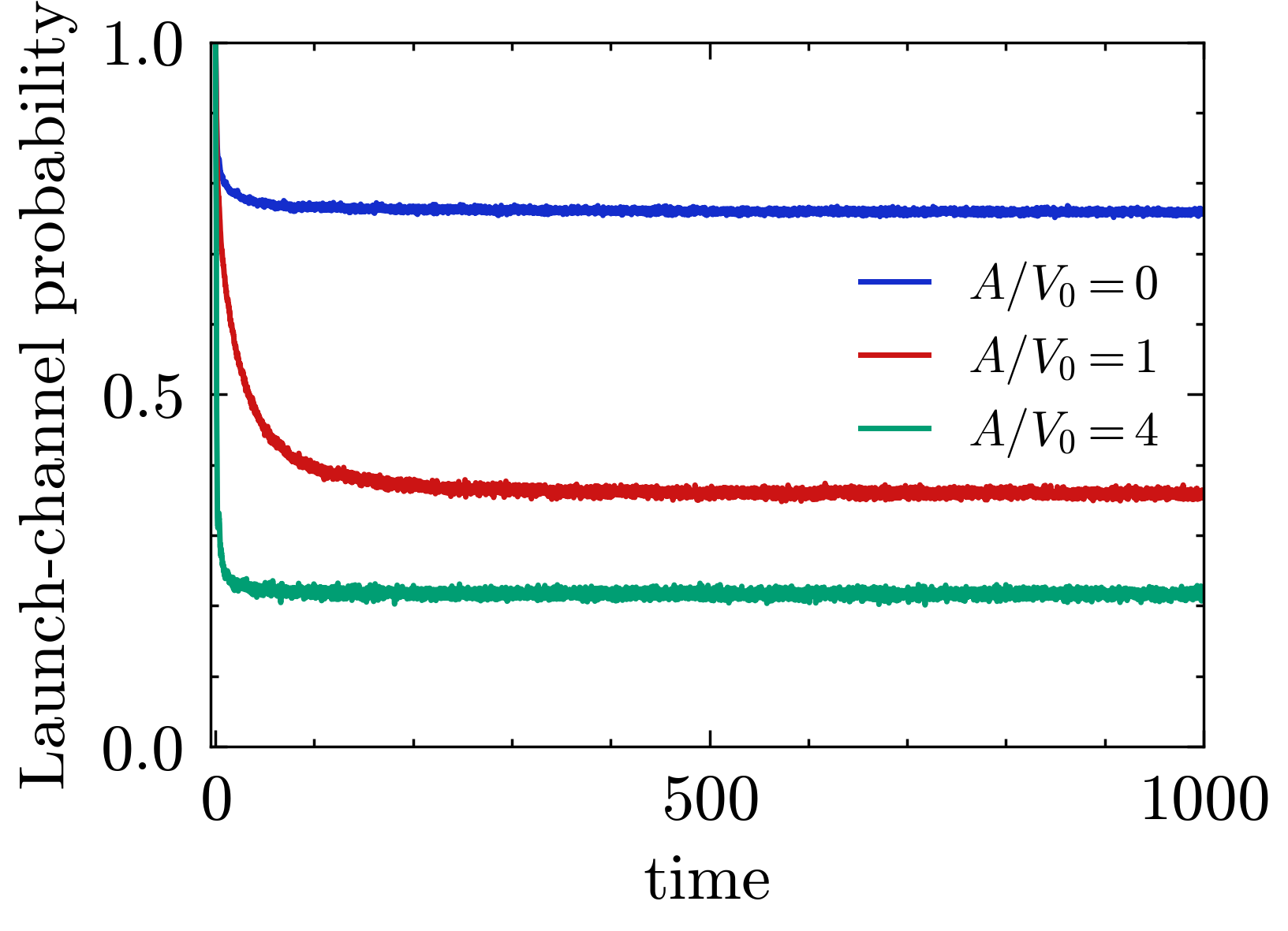}
    \caption{Classical launch-channel probability for $A/V_0=0,1,$ and $4$, with the impurity centered in the launch-channel. The probability is the fraction of ensemble trajectories within that channel at each time. Increasing impurity strength reduces classical confinement over the interval studied.}
\label{fig:classical_channel_probability}
\end{figure}

Figure~\ref{fig:classical_channel_probability} shows the launch-channel probability for classical ensembles initialized from the Gaussian phase-space distribution corresponding to the initial quantum wavepacket. After the initial transient, the probabilities remain near $0.76$, $0.36$, and $0.22$ for $A/V_0=0,1,$ and $4$, respectively. Thus, the impurity weakens classical confinement, in contrast to the quantum enhancement at intermediate strength. This comparison includes the initial ensemble's spread in position, momentum, and energy, complementing the Poincar\'e sections evaluated at the mean launch energies.
\section{Finite-Time Channel Average}
\label{sec:finite_time_derivation}

For a fixed impurity amplitude $A$ and launch-channel $\alpha_0$, write
\begin{equation}
H(A)|\phi_n(A)\rangle
=
E_n(A)|\phi_n(A)\rangle,
\qquad
c_n(A)=\langle\phi_n(A)|\psi_{\alpha_0}(0)\rangle.
\label{eq:sm_eigenvalue_equation}
\end{equation}
Here, $c_n(A)$ abbreviates $c_n^{(\alpha_0)}(A)$. The retained initial state and its evolution are
\begin{equation}
\begin{aligned}
|\psi_{\alpha_0}^{(N)}(0;A)\rangle
&=
\sum_{n=1}^{N}c_n(A)|\phi_n(A)\rangle,\\
|\psi_{\alpha_0}^{(N)}(t;A)\rangle
&=
\sum_{n=1}^{N}
c_n(A)e^{-iE_n(A)t}|\phi_n(A)\rangle.
\end{aligned}
\label{eq:sm_time_evolution}
\end{equation}
All expressions below are exact within this retained expansion. We suppress its superscript $(N)$ on the channel probabilities for readability.

With
\begin{equation}
P_{nm}^{(\alpha_0)}(A)
=
\langle\phi_n(A)|\hat P_{\alpha_0}|\phi_m(A)\rangle,
\label{eq:sm_channel_projector_matrix}
\end{equation}
the launch-channel probability is
\begin{equation}
\mathcal P_{\alpha_0}(t;A)
=
\sum_{n,m=1}^{N}
c_n^*(A)c_m(A)P_{nm}^{(\alpha_0)}(A)
e^{i\Delta E_{nm}(A)t},
\label{eq:sm_time_dependent_channel_probability}
\end{equation}
where $\Delta E_{nm}(A)=E_n(A)-E_m(A)$. Its average over $0\leq t\leq T$ is
\begin{equation}
\begin{aligned}
\overline{\mathcal P}_{\alpha_0}^{(T)}(A)
&=
\frac{1}{T}\int_0^T\mathcal P_{\alpha_0}(t;A)\,dt\\
&=
\sum_{n,m=1}^{N}
c_n^*(A)c_m(A)P_{nm}^{(\alpha_0)}(A)
F_T\!\left(\Delta E_{nm}(A)\right),
\end{aligned}
\label{eq:sm_finite_time_average_kernel}
\end{equation}
with
\begin{equation}
F_T(\Delta E)
=
\frac{1}{T}\int_0^T e^{i\Delta E t}\,dt.
\label{eq:sm_time_kernel_definition}
\end{equation}
For $\Delta E\neq0$,
\begin{equation}
F_T(\Delta E)
=
\frac{e^{i\Delta E T}-1}{i\Delta E T}
=
e^{i\Delta E T/2}
\operatorname{sinc}\!\left(\frac{\Delta E T}{2}\right),
\label{eq:sm_time_kernel_sinc}
\end{equation}
where $\operatorname{sinc}(z)=\sin(z)/z$ and $\operatorname{sinc}(0)=1$. The continuous limit gives $F_T(0)=1$, also obtained directly from Eq.~\eqref{eq:sm_time_kernel_definition}.

The result is real because terms with indices $(n,m)$ and $(m,n)$ are complex conjugates. Pairs with $|\Delta E_{nm}|T\lesssim1$ are not strongly suppressed by averaging and can contribute appreciably when the corresponding wavepacket-overlap coefficients and channel projector matrix elements are significant.

\subsection{Infinite-time limit and degeneracy}

For any fixed nonzero gap,
$F_T(\Delta E)\rightarrow0$ as $T\rightarrow\infty$,
whereas $F_T(0)=1$. The infinite-time mean within the retained expansion is therefore
\begin{equation}
\overline{\mathcal P}_{\alpha_0}^{(\infty)}(A)
=
\sum_{\substack{n,m=1\\E_n(A)=E_m(A)}}^{N}
c_n^*(A)c_m(A)P_{nm}^{(\alpha_0)}(A).
\label{eq:sm_infinite_time_degenerate_average}
\end{equation}
For a nondegenerate spectrum, this reduces to
\begin{equation}
\overline{\mathcal P}_{\alpha_0}^{(\infty)}(A)
=
\mathcal P_{\alpha_0}^{\mathrm{diag}}(A)
=
\sum_{n=1}^{N}|c_n(A)|^2p_{\alpha_0n}(A),
\label{eq:sm_diagonal_ensemble}
\end{equation}
where $p_{\alpha_0n}=P_{nn}^{(\alpha_0)}$.
Exactly degenerate states can retain off-diagonal contributions. Within such subspaces, the diagonal term alone can depend on the eigenbasis, whereas the complete mean is basis independent. We retain the full finite-time expression throughout.

\subsection{Comparison with direct propagation}

\begin{figure}[h]
    \centering
    \includegraphics[width=0.45\linewidth]
    {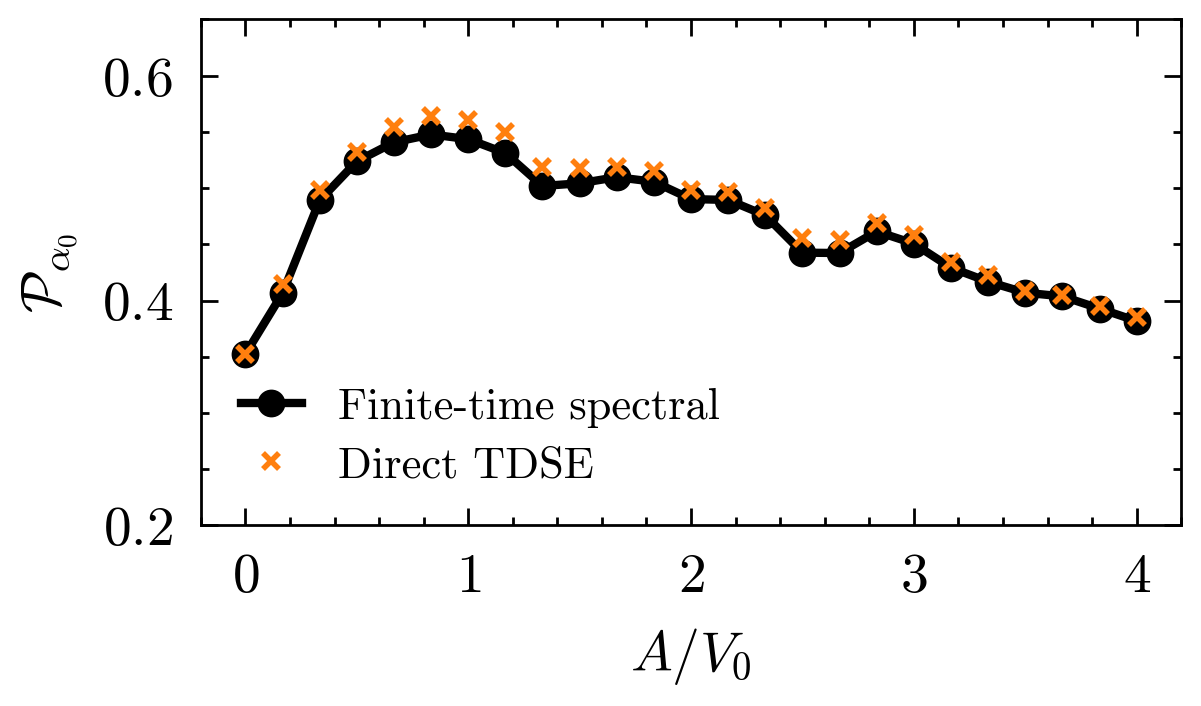}
    \caption{Launch-channel probability averaged over $0\leq t\leq1000$ versus impurity strength $A/V_0$. The finite-time spectral result, including diagonal
    and off-diagonal terms (black circles), agrees closely with direct split-operator propagation (orange crosses).}
    \label{fig:tdse-finiteTimeAve}
\end{figure}

Figure~\ref{fig:tdse-finiteTimeAve} compares the spectral average for $T=1000$ with the time average obtained from direct split-operator propagation. Their close agreement
across the impurity-strength scan supports the numerical accuracy of the retained spectral evaluation, including its off-diagonal contributions.

\section{Cumulative Channel Coherence}

To identify the energy gaps contributing to the mean interference correction, we define the cumulative signed coherence
\begin{equation}
\begin{aligned}
K_{\alpha_0}(\varepsilon;T)
&=
2\,\mathrm{Re}
\sum_{\substack{1\leq n<m\leq N\\
|E_n-E_m|\leq\varepsilon}}
c_n^{(\alpha_0)*}c_m^{(\alpha_0)}
P_{nm}^{(\alpha_0)}\\
&\qquad\times
e^{i(E_n-E_m)T/2}
\operatorname{sinc}\!\left[
\frac{(E_n-E_m)T}{2}
\right],
\end{aligned}
\label{eq:cumulative_channel_coherence}
\end{equation}
where $\varepsilon$ is an energy-gap cutoff and the dependence on $A$ is implicit. Increasing $\varepsilon$ includes additional pairs, whose signed contributions can increase or decrease the sum. Once all retained pairs are included,
\begin{equation}
K_{\alpha_0}(\infty;T)
=
\overline{\mathcal C}_{\alpha_0}^{(T)}.
\label{eq:cumulative_coherence_limit}
\end{equation}

\begin{figure}[h!]
    \centering
    \includegraphics[width=0.45\linewidth]
    {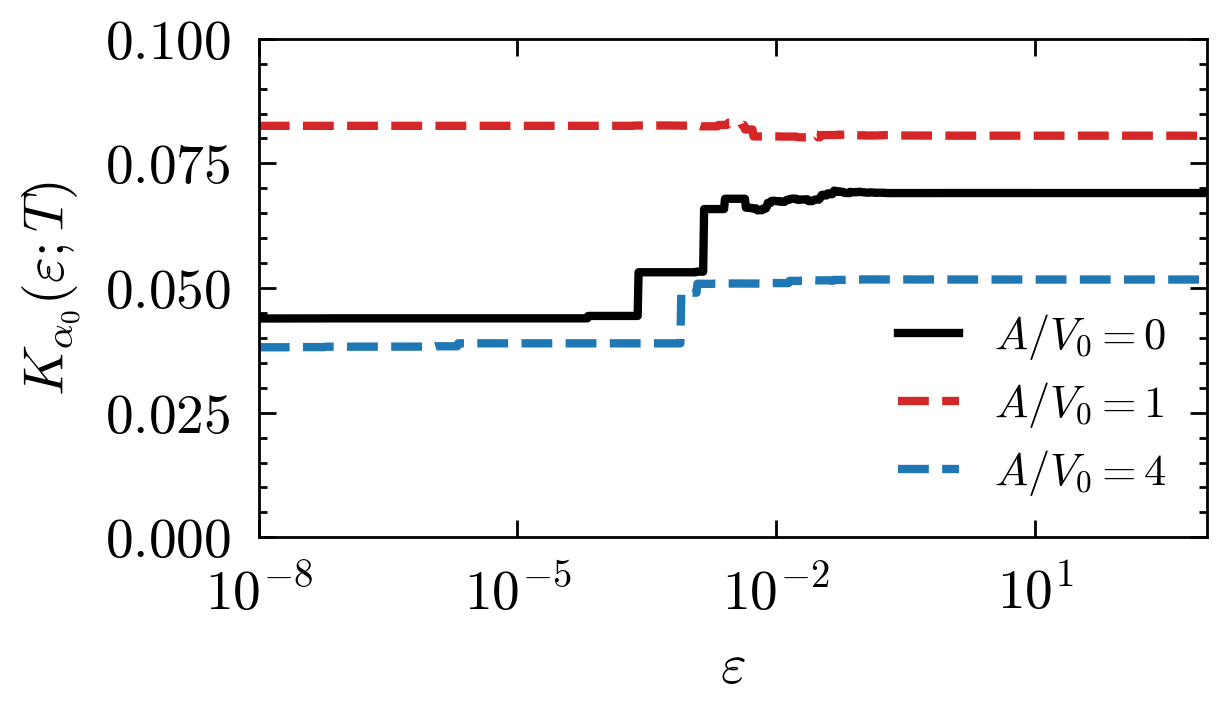}
    \caption{Cumulative signed coherence
    $K_{\alpha_0}(\varepsilon;T)$ for $T=1000$ and
    $A/V_0=0,1,$ and $4$, including eigenstate pairs with
    $|E_n-E_m|\leq\varepsilon$. At the largest cutoffs,
    each curve approaches the full time-averaged
    interference correction.}
    \label{fig:signed_coherence}
\end{figure}

Figure~\ref{fig:signed_coherence} shows that small-gap pairs dominate the net correction; the smallest plotted cutoff already includes all smaller gaps and any exact degeneracies. Larger gaps increase the correction overall for $A/V_0=0$ and $4$ but slightly decrease it for $A/V_0=1$, reflecting cancellation between signed interference terms. The plateaus indicate negligible additional net contributions, not vanishing individual terms. Pairs with $|E_n-E_m|T\ll1$ retain nearly constant relative phases, while rapidly oscillating terms tend to average out but can still produce instantaneous fluctuations without loss of coherence. Over the interval studied, interference adds a positive mean correction, while the diagonal contribution captures the main nonmonotonic confinement trend.

\section{Dependence on Impurity Position and launch-channel}

\subsection{Off-center impurities}

\begin{figure}[h]
    \centering
    \includegraphics[width=0.45\linewidth]
    {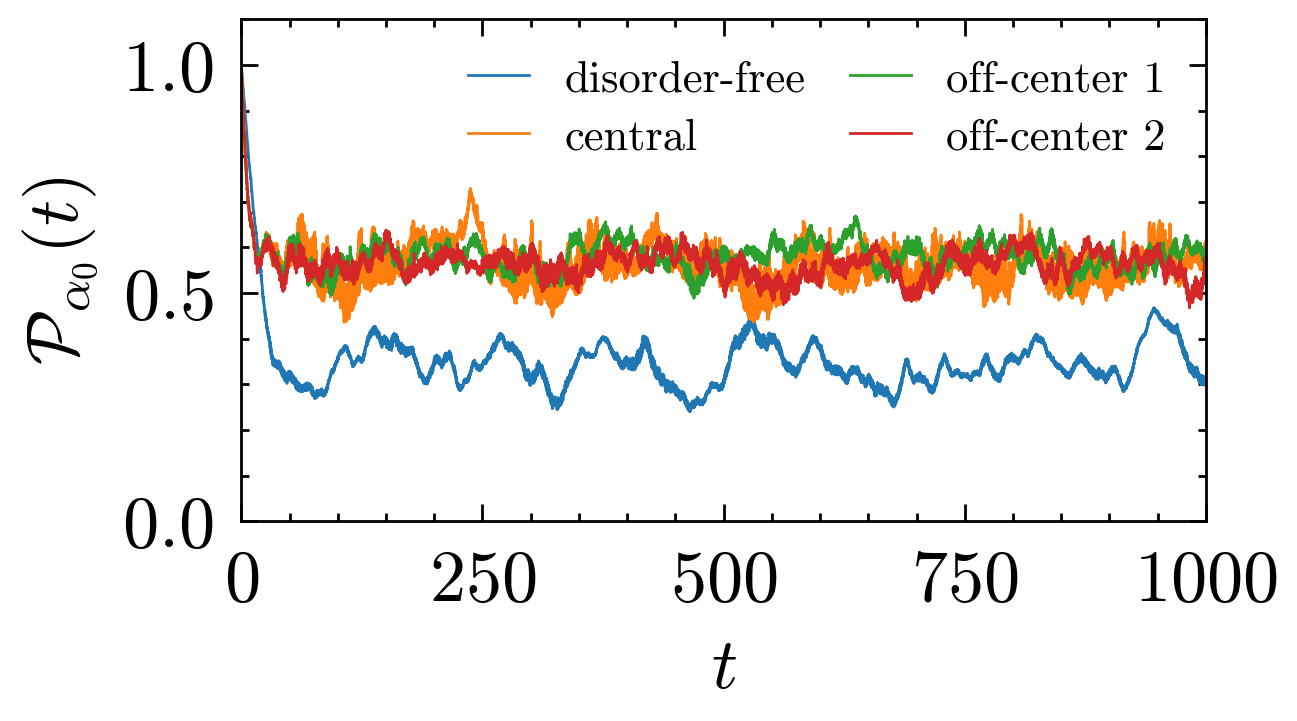}
    \caption{Comparison of channel confinement for the
    impurity positions indicated in the figure, at fixed
    impurity amplitude and width. The off-center
    configurations yield confinement comparable to
    central placement.}
    \label{fig:off-center}
\end{figure}

The off-center impurity configurations examined yield
comparable confinement at the same amplitude and width
(Fig.~\ref{fig:off-center}). This comparison supports the
central placement used in the main text as a representative
configuration.
\subsection{Different launch-channels}

\begin{figure}[h]
    \centering
    \includegraphics[width=0.45\linewidth]
    {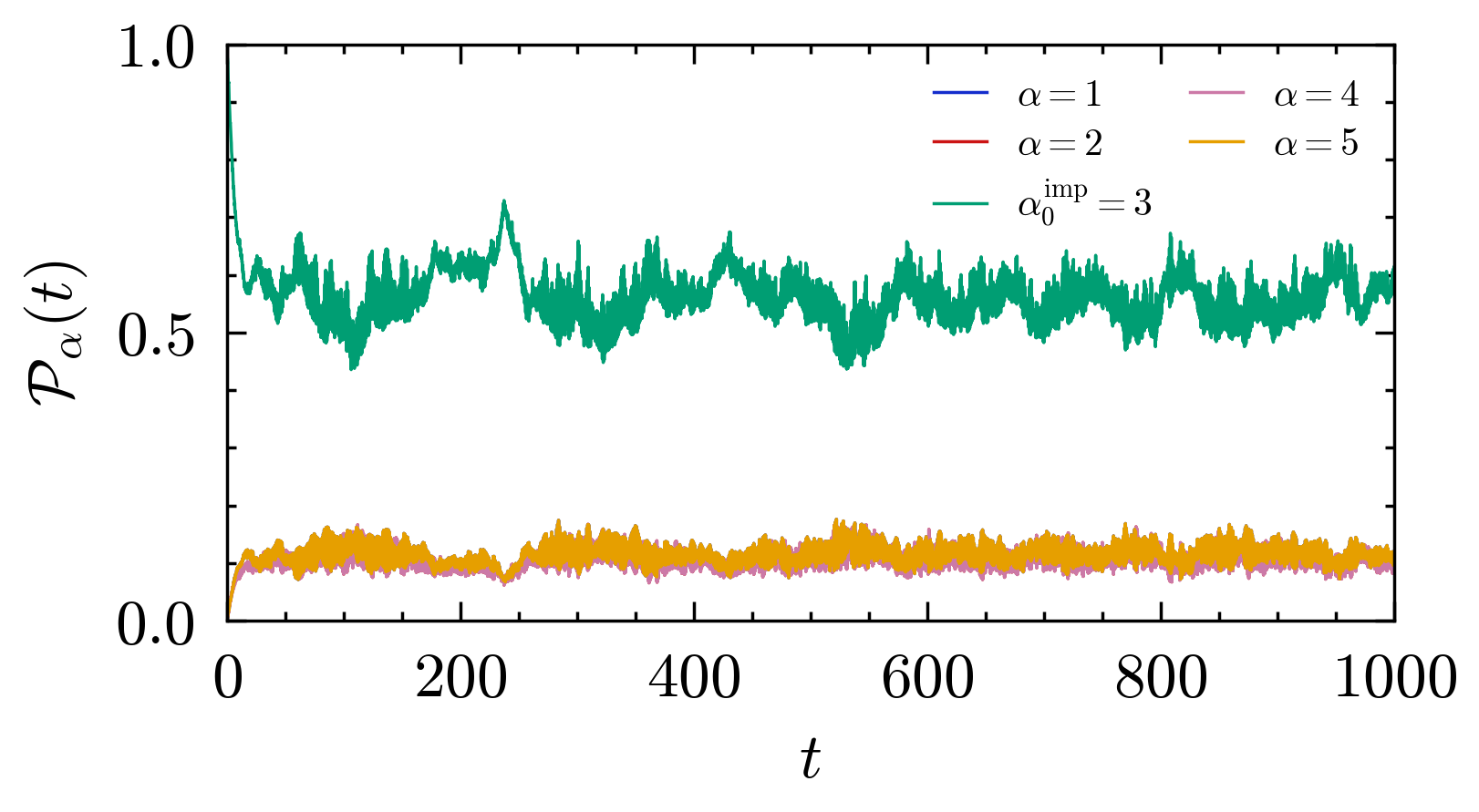}
    \hfill
    \includegraphics[width=0.45\linewidth]
    {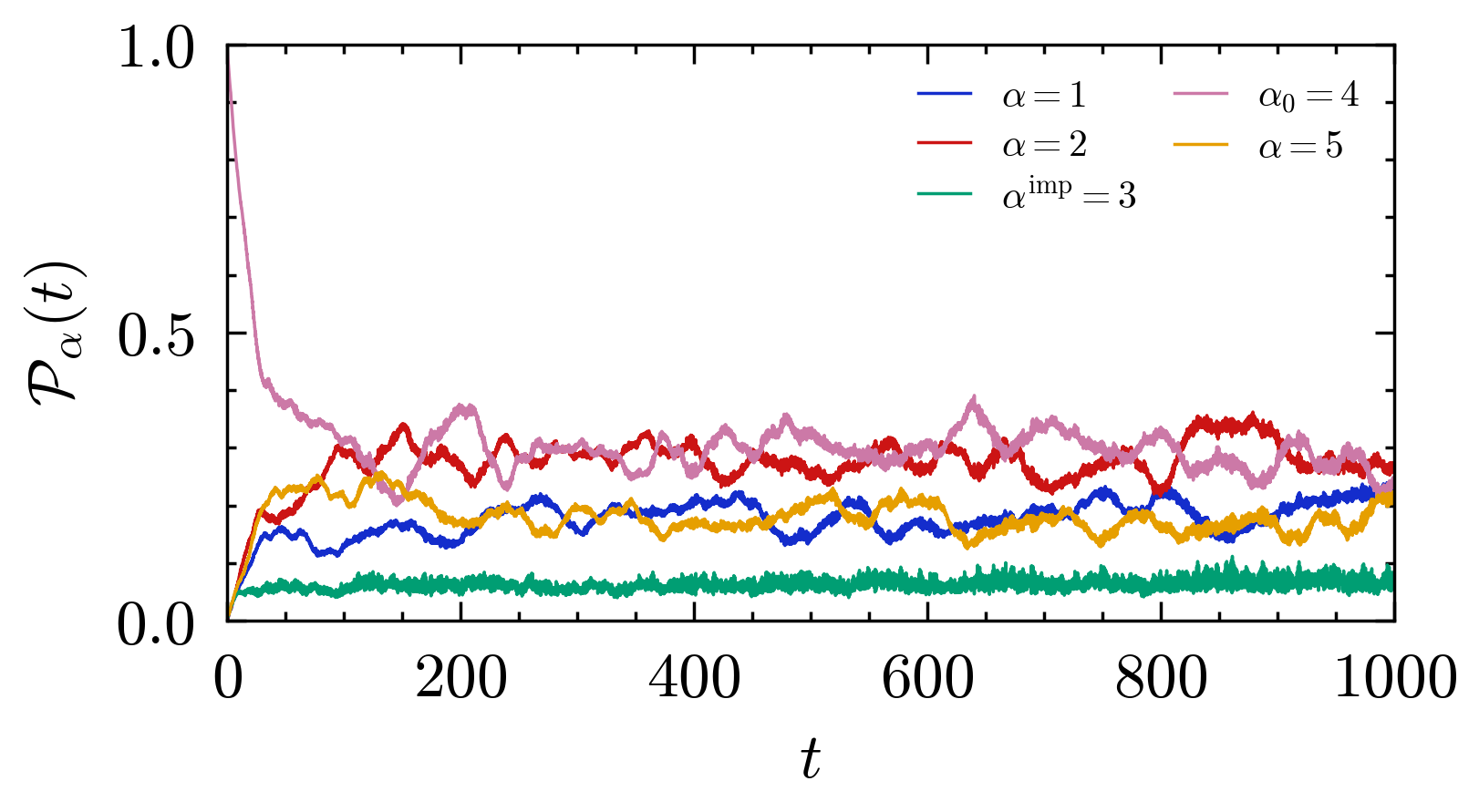}
    \caption{Channel probabilities $\mathcal P_\alpha(t)$
    for an impurity in channel $\alpha^{\mathrm{imp}}=3$.
    Left: launching in the impurity channel ($\alpha_0=3$)
    produces enhanced confinement. Right: launching in
    a neighboring channel ($\alpha_0=4$) redistributes
    probability among the unperturbed channels, while
    the impurity channel remains the least populated
    after the initial transient.}
    \label{fig:diffLaunch}
\end{figure}

Keeping the impurity in channel $\alpha^{\mathrm{imp}}=3$, we compare launches in that channel and in the neighboring channel $\alpha_0=4$ (Fig.~\ref{fig:diffLaunch}). In the
first case, the launch-channel remains predominantly populated. In the second, probability spreads among the unperturbed channels, while the impurity channel remains
the least populated after the initial transient. Changing the initial state changes its eigenstate overlaps without changing the Hamiltonian or channel projector
matrix elements. The observed redistribution is consistent with weaker overlap with states concentrated in the impurity channel and stronger overlap with states extending across the unperturbed channels. Thus, the same impurity can enhance confinement for a packet launched in its channel while limiting population of that channel for a packet launched elsewhere.

\end{document}